\documentclass[11pt]{article}

\usepackage{a4}
\usepackage{amsthm}
\usepackage[english]{babel}
\usepackage{german}
\usepackage{graphicx}
\usepackage{subfigure}
\usepackage{amsmath}
\usepackage{array}
\usepackage{setspace}
\usepackage{enumerate}

\newtheorem{thm}{Theorem}
\theoremstyle{definition}
\newtheorem{dfn}{Definition}
\theoremstyle{remark}

\theoremstyle{plain}
\newtheorem{lem}{Lemma}

\begin{document}

\title{A polynomial algorithm for the k-cluster problem \\
on interval graphs}
\author{George B. Mertzios \\
Department of Computer Science,\\
RWTH Aachen University\\
\underline{\emph{mertzios@cs.rwth-aachen.de}}}
\maketitle

\begin{abstract}
This paper deals with the problem of finding, for a given graph and a given
natural number k, a subgraph of k nodes with a maximum number of edges. This
problem is known as the \emph{k-cluster problem} and it is NP-hard on
general graphs as well as on chordal graphs. In this paper, it is shown that
the k-cluster problem is solvable in polynomial time on interval graphs. In
particular, we present two polynomial time algorithms for the class of
proper interval graphs and the class of general interval graphs,
respectively. Both algorithms are based on a matrix representation for
interval graphs. In contrast to representations used in most of the previous
work, this matrix representation does not make use of the maximal cliques in
the investigated graph.\newline
\end{abstract}

\setcounter{page}{1} \setcounter{tocdepth}{4} \setcounter{secnumdepth}{4} 
\fboxsep 0mm \setlength{\baselineskip}{3ex}

\section{Introduction}

A graph $G$ is called an \emph{interval graph }if its nodes can be assigned
to intervals on the real line so that two nodes are adjacent in $G$ if and
only if their assigned intervals intersect. The set of intervals assigned to
the nodes of $G$ is called a \emph{realization }of $G$. A \emph{proper
interval graph} is an interval graph that has an intersection model, in
which no interval contains another one strictly. Interval and proper
interval graphs have been studied extensively in the literature and several
linear-time algorithms are known for their recognition \cite%
{Hsu93,Corneil98,Corneil95}. They are important for their applications to
scheduling problems, biology, VLSI circuit design, as well as to psychology
and social sciences in general \cite{Golumbic04,Carrano88}.

The class of interval graphs is of major importance, while studying the
complexity of several difficult optimization problems, which are solvable in
polynomial time on them, but NP-hard in the general case. Some of these
problems are the maximum clique \cite{Gupta82}, the maximum independent set 
\cite{Gupta82,Ju92}, the Hamiltonian cycle and the Hamiltonian path \cite%
{Chang93}.

This paper deals with the problem of finding, for a given graph and a given
natural number k, a subgraph on k nodes and of maximum number of edges. This
problem is called the \emph{k-cluster problem}. Until now it is known that
the k-cluster problem is NP-hard as a generalization of the maximum clique
problem. It remains NP-hard, even when restricted to comparability graphs,
as well as on bipartite graphs and chordal graphs \cite{Corneil84}. On the
other side, it has been proved that there are polynomial algorithms for the
k-cluster problem on cographs, as well as on $k$-trees and split graphs \cite%
{Corneil84}. Furthermore, it has been proved that the decision version of
the k-cluster problem is solvable in polynomial time, when searching for
fixed-density $k$-subgraphs, while it remains NP-hard, when searching for a $%
k$-subgraph with density at least $f\left( k\right) =\Omega \left( {%
k^{\varepsilon }}\right) $ edges, for some $\varepsilon >0$ \cite%
{Holzapfel03}. Finally, there are also some other polynomial time algorithms
designed for the k-cluster problem on some special classes of the proper
interval graphs, e.g., of the graphs, whose clique graph is a simple path 
\cite{Liazi05}.

In the present work, it is proved that the k-cluster problem on proper
interval graphs, as well as on the general class of interval graphs, is
solvable in polynomial time and thus the corresponding open problem stated
in \cite{Corneil84} is answered. To this end, a matrix representation, which
characterizes these classes of graphs, is used here. This representation
does not use their maximal cliques, as the vast variety of the existing
characterizations do.

\section{The interval graphs in the general case}

\label{Int_Gen} Without loss of generality, we may suppose that all
intervals in a realization of an interval graph are closed, i.e. of the form 
$\left[ {a,b}\right] $. However, this representation is too general. To this
end, a more suitable interval representation form is presented in Definition~%
\ref{Def_1} \cite{Mertzios07}. Recall that an interval graph can be
recognized in linear time \cite{Hsu93,Corneil98}. In the following, suppose
we are given a realization of an interval graph $G$ on $n$ nodes.

\begin{dfn}
\label{Def_1} A representation of $n$ intervals, having the following
properties, is called a \emph{Normal Interval Representation (NIR) form}:

\begin{enumerate}
\item \vspace{-0.3cm} all intervals are of the form $\left[ {i,j} \right)$,
where $0 \le i<j \le n$,

\item \vspace{-0.3cm} exactly one interval begins at $i$, for every $i\in
\left\{ {0,1,...,n-1} \right\}$.
\end{enumerate}
\end{dfn}

Suppose we are given a realization of the interval graph $G$. It can be
converted to another realization of the same graph, in which all $2n$
endpoints are distinct in the real line. This can be done simply by
disturbing them sufficiently, so that the structure of the graph remains
unchanged, under the condition that the relative order of the left endpoints
of any two intervals is not being reversed. After that, the arbitrary closed
interval $\left[ {a,b}\right] $ may be replaced by $\left[ {a,b}\right) $,
since the intersection of any two intervals, if such occurs, is a
non-trivial interval. In the sequel, any interval's right endpoint may be
moved to the next greater interval's left endpoint in the current
realization, resulting thus in exactly $n+1$ distinct endpoints altogether.
Finally, all these endpoints may be moved bijectively to the points $%
0,1,...,n$, obtaining thus an NIR form of $G$ in linear time $O(n)$.

\begin{lem}
\label{Lem_2} An arbitrary graph is an interval graph iff it can be
represented by the NIR form.
\end{lem}

\begin{proof}
An NIR form is clearly a set of intervals and thus it corresponds to an interval graph. 
Conversely, since any interval graph can be represented by an NIR form, this representation holds as a characterization 
of interval graphs.
\end{proof}
Since no two intervals in the NIR form share a common left endpoint, it is
possible to define a perfect order over them. Let the $i^{th} $ interval be $%
\left[ {i-1,b}\right) $. Now recall the Heaviside function: 
\begin{equation*}
H\left( x\right) :=%
\begin{cases}
1, & \text{if }x\geq 0 \\ 
0, & \text{otherwise}%
\end{cases}%
\end{equation*}

\begin{dfn}
\label{Def_2} Consider the $i^{th}$ interval $\left[ {i-1,b} \right)$ of the
NIR form of the interval graph $G$, for which we define the quantity $x_i
:=b-i$. Then, the square matrix 
\begin{equation*}  \label{eq2}
H_G \left( {i,j} \right):= 
\begin{cases}
H\left( {x_j +j-i} \right), & \text{if } i>j \\ 
0, & \text{otherwise}%
\end{cases}%
\end{equation*}
is called the \emph{Normal Interval Representation (NIR) matrix} of $G$.
\end{dfn}

In the above definition the quantity $x_{i}$ equals the number of intervals
among the $\left( {i+1}\right) ^{th},...,n^{th}$ ones that intersect with
the $i^{th}$ one. $H_{G}$ is a lower triangular matrix with zero diagonal,
having a chain of $x_{i}$ consecutive $1$'s under the $i^{th}$ diagonal
element and all the remaining matrix entries being zero. It can be seen also
as the lower triangular portion of the adjacency matrix of $G$, where
however rows and columns are ordered in a particular way. Specifically, the $%
i^{th}$ interval of $G$ is represented schematically by the $i^{th}$ column
of $H_{G}$. Figure~\ref{Fig_1a} shows an example of the form of $H_{G}$.

Denote further the desired $k$-subgraph of $G$ with the maximum number of
edges as $C_{k}$. Join the variable $z_{i}\in \left\{ {0,1}\right\} $ to the 
$i^{th}$ interval. The case $z_{i}=1$ indicates that the $i^{th}$ node of $G$%
, i.e. the $i^{th}$ interval of its NIR form, is included in $C_{k}$. Let
now $1\leq j<i\leq n$. The $j^{th}$ and the $i^{th}$ intervals intersect in $%
C_{k}$ if and only if the quantity $z_{j}\cdot z_{i}\cdot H\left( {x_{j}+j-i}%
\right) \in \left\{ {0,1}\right\} $ equals one. Indeed, in this case both
intervals have been chosen in $C_{k}$, i.e. $z_{i}=z_{j}=1$ and,
simultaneously, the $j^{th}$ interval ends strictly further than $i-1$,
where the $i^{th}$ one begins, i.e. $H\left( {x_{j}+j-i}\right) =1$. Thus,
the number of intersections among the $k$ intervals of the realization of $%
C_{k}$ equals 
\begin{equation}
\sum\nolimits_{i=2}^{n}{\sum\nolimits_{j=1}^{i-1}{z_{j}\cdot z_{i}\cdot
H\left( {x_{j}+j-i}\right) }}=z^{T}\cdot H_{G}\cdot z  \label{eq1}
\end{equation}%
where $z=\left[ {{%
\begin{array}{*{20}c}
 {z_1 } \hfill & {z_2 } \hfill & \cdots \hfill & {z_n } \hfill \\
\end{array}}}\right] ^{T}$ and $H_{G}$ is the NIR matrix of $G$.

Since $C_{k}$ has exactly $k$ nodes, exactly $k$ entries of the vector $z$
are one. Thus, the k-cluster problem on $G$ is equivalent to finding the
appropriate subset $I\subseteq \left\{ {1,2,...,n}\right\} $ of the
satisfied entries of $z$, with $\left\vert I\right\vert =k$, so that the
following quantity is maximized: 
\begin{equation}
\sum\nolimits_{\mathop {i,j\in I}\limits_{i>j}}{H_{G}\left( {i,j}\right) }%
=\sum\nolimits_{\mathop {i,j\in I}\limits_{i>j}}{H\left( {x_{j}+j-i}\right) }
\label{eq3}
\end{equation}

\begin{lem}
\label{Lem_3} Any maximal clique of $G$ corresponds bijectively to a row of
its NIR matrix $H_G $, in which at least one of its unit elements or its
zero diagonal element does not have any chain of $1$'s below it.
\end{lem}

\begin{proof}
Consider an arbitrary row of $H_{G}$, let it be the $i^{th}$ one, in which
exactly the $i_{1}^{th},i_{2}^{th},...,i_{r}^{th}$ elements equal one.
Clearly, the $i^{th}$ and the $j^{th}$ intervals intersect for every $j\in
\left\{ {i_{1},i_{2},...,i_{r}}\right\} $, since $H_{G}\left( {i,j}\right) =1
$. The $i_{1}^{th},i_{2}^{th},...,i_{r}^{th}$ intervals of $G$ intersect
each other also, due to the NIR form of $H_{G}$. Thus, the $%
i_{1}^{th},i_{2}^{th},...,i_{r}^{th},i^{th}$ intervals build a clique $Q$ in 
$G$. Consider now the case that in this row at least one of its $%
i_{1}^{th},i_{2}^{th},...,i_{r}^{th},i^{th}$ elements, say the $j^{th}$ one,
does not have any chain of $1$'s below it. Suppose also that there exists
another clique $Q^{\prime }$ in $G$, which strictly includes $Q$. Since $%
H_{G}\left( {\ell _{1},j}\right) =H_{G}\left( {i,\ell _{2}}\right) =0$ for
every $\ell _{1}>i$ and $\ell _{2}\in \left\{ {1,2,...,i}\right\} \setminus
\left\{ {i_{1},i_{2},...,i_{r}}\right\} $, the $\ell _{1}^{th}$ and the $%
j^{th}$, as well as the $i^{th}$ and the $\ell _{2}^{th}$ intervals, do not
intersect. Therefore, $Q^{\prime }$ can not be a clique, which is a
contradiction. Thus, $Q$ is a maximal clique.

Conversely, let $Q$ be a maximal clique in $G$, which contains the $i_1
^{th},i_2 ^{th},...,i_{\left| Q \right|} ^{th}$ intervals of its NIR form,
where $i_1 < i_2 < ... < i_{\left| Q \right|} $. Consider now the $i_{\left|
Q \right|} ^{th}$ row of $H_G $. Since $Q$ is a clique, the $i_1 ^{th},i_2
^{th},...,i_{\left| Q \right| - 1} ^{th}$ intervals intersect with the $%
i_{\left| Q \right|} ^{th}$ one and therefore $H_G \left( {i_{\left| Q
\right|} ,j} \right) = 1$ for every $j \in \left\{ {i_1 ,i_2 ,...,i_{\left|
Q \right| - 1} } \right\}$. Suppose $i_{\left| Q \right|} < n$. Then, if $%
H_G \left( {i_{\left| Q \right|} + 1,j} \right) = 1$ for every $j \in
\left\{ {i_1 ,i_2 ,...,i_{\left| Q \right|} } \right\}$, the $i_{\left| Q
\right| + 1} ^{th}$ row corresponds to another clique $Q^{\prime }$ that
includes $Q$ strictly, which is a contradiction. Thus, at least one of the $%
i_1 ^{th},i_2 ^{th},...,i_{\left| Q \right|} ^{th}$ elements of the $%
i_{\left| Q \right|} ^{th}$ row does not have any chain of $1$'s below it.
Finally, in the case where $i_{\left| Q \right|} = n$, obviously none of the 
$i_1 ^{th},i_2 ^{th},...,i_{\left| Q \right|} ^{th}$ elements of the $%
i_{\left| Q \right|} ^{th}$ has any chain of $1$'s below it.
\end{proof}

\section{The proper interval graph case}

\label{Pr_Case} Consider now the case that $G$ is a proper interval graph.
Since $G$ is also an interval graph, it can be represented by the NIR form,
which however has an additional property, as described in Definition~\ref%
{Def_3}.

\begin{dfn}
\label{Def_3}
An NIR form of $n$ intervals is called a \emph{Stair Normal Interval Representation (SNIR) form}, iff it has 
the following additional property:
\begin{enumerate} [\hspace{0.5cm} ]
\item \vspace{-0.2cm} If for the intervals $\left[ {a,b} \right)$ and $\left[ {c,d} \right)$, $a<c$ holds, 
then $b\le d$ also holds.
\end{enumerate}
\end{dfn}

\begin{lem}
\label{Lem_4} An arbitrary proper interval graph $G$ can be converted to the
SNIR form.
\end{lem}

\begin{proof}
Suppose we are given an arbitrary realization of $G$, in which no interval contains another strictly. 
Consider the case that in this realization the left endpoint of the interval $v_1 =\left[ {a,b} \right]$ 
is strictly less than the left endpoint of the interval $v_2 =\left[ {c,d} \right]$, i.e., $a<c$. 
Then the same also do their right endpoints respectively. i.e., $b<d$, since otherwise $v_2 $ would strictly 
include $v_1 $, which is a contradiction. Since $G$ is also an interval graph, it can be converted to the 
NIR form, as described above. Suppose that $v_1 $ and $v_2 $ are converted to the intervals 
$v_1 '=\left[ {a',b'} \right)$ and $v_2 '=\left[ {c',d'} \right)$ in the resulting NIR form respectively. 
Then, $a'<c'$ holds, since the relative order of the interval left points $a$ and $c$ is not being 
reversed during the conversion of $G$ to the NIR form; also $b'\le d'$ holds, since the right endpoints $b$ and $d$ 
may be ``aligned'' by the left interval endpoints of the graph. Thus, the obtained NIR form satisfies the condition 
of Definition~\ref{Def_3}, i.e., it is an SNIR form. Note that in the special case of two initially identical intervals, 
i.e., $a=c$ and $b=d$, we obtain the same right endpoints $b'=d'$ for them in the resulting NIR form, while their 
left endpoints are ordered by increasing order, i.e., in this case the obtained NIR form is also an SNIR form.
\end{proof}

\begin{dfn}
\label{Def_4} The NIR matrix $H_G$ that corresponds to the SNIR form of a
proper interval graph $G$ is called the \emph{Stair Normal Interval
Representation (SNIR) matrix} of $G$.
\end{dfn}

\begin{dfn}
\label{Def_5} Consider the SNIR matrix $H_G$ of the proper interval graph $G$%
. The matrix element $H_G(i,j)$ is called a \emph{pick} of $H_G$, iff:

\begin{enumerate}
\item \vspace{-0.2cm} $i\ge j$,

\item \vspace{-0.2cm} if $i>j$ then $H_{G}(i,j)=1$,

\item \vspace{-0.3cm} $H_G(i,k)=0$, for every $k\in \left\{ {1,2,...,j-1}
\right\}$ and

\item \vspace{-0.3cm} $H_{G}(\ell ,j)=0$, for every $\ell \in \left\{ {%
i+1,i+2,...,n}\right\} $.
\end{enumerate}

\vspace{-0.2cm} Given the pick $H_{G}(i,j)$ of $H_{G}$, the set 
\begin{equation*}
S:=\left\{ {H_{G}\left( {k,\ell }\right) :i\geq k\geq \ell \geq j}\right\}
\end{equation*}%
of matrix entries is called the \emph{stair} of $H_{G}$, which corresponds
to this particular pick.
\end{dfn}

\begin{figure}[htb]
\centering
\mbox{
      \subfigure[]{        
        \label{Fig_1a}
        \includegraphics[width=5.8cm]{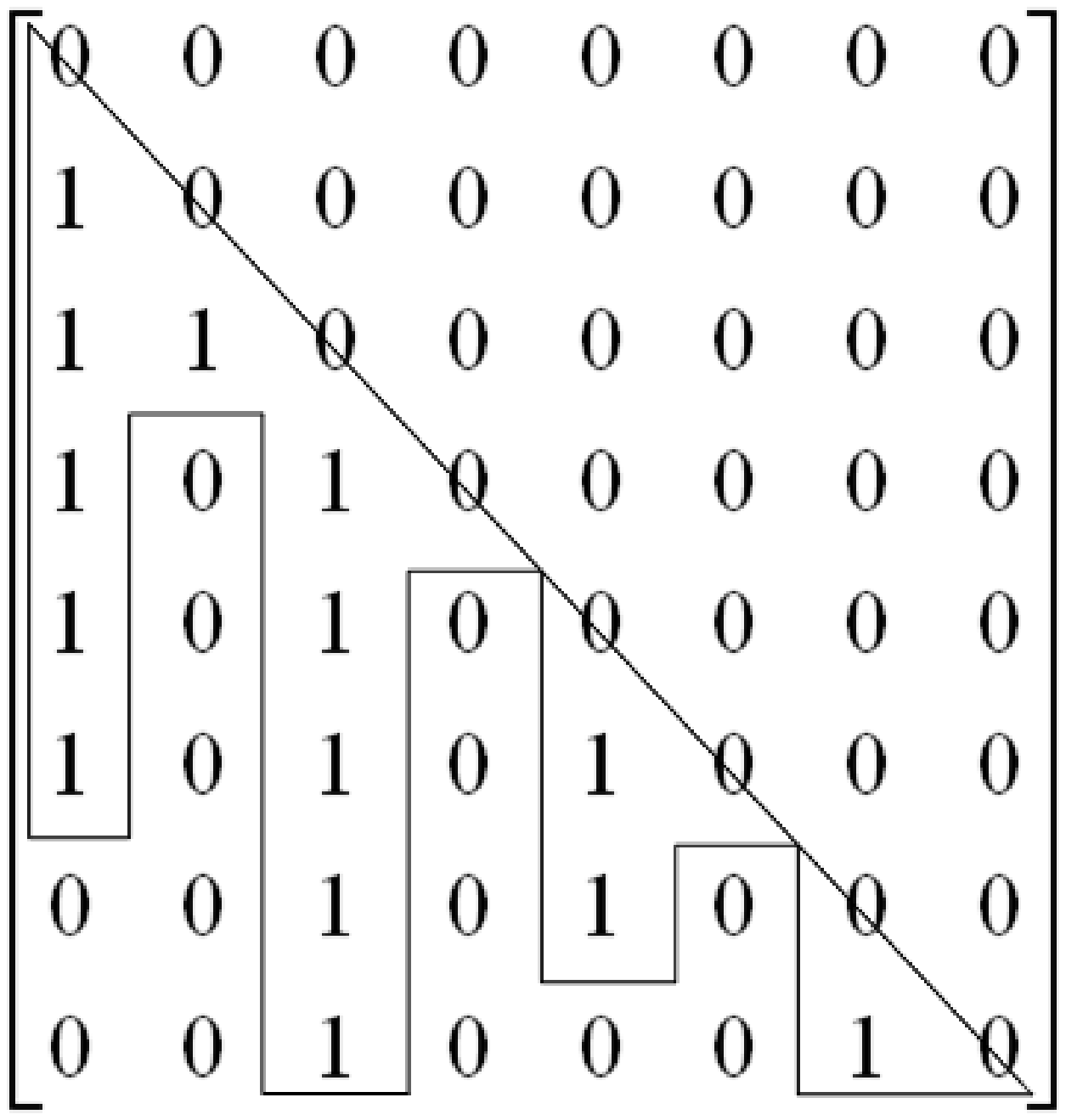}} 
        
      \subfigure[]{
        \label{Fig_1b}
        \includegraphics[width=5.8cm]{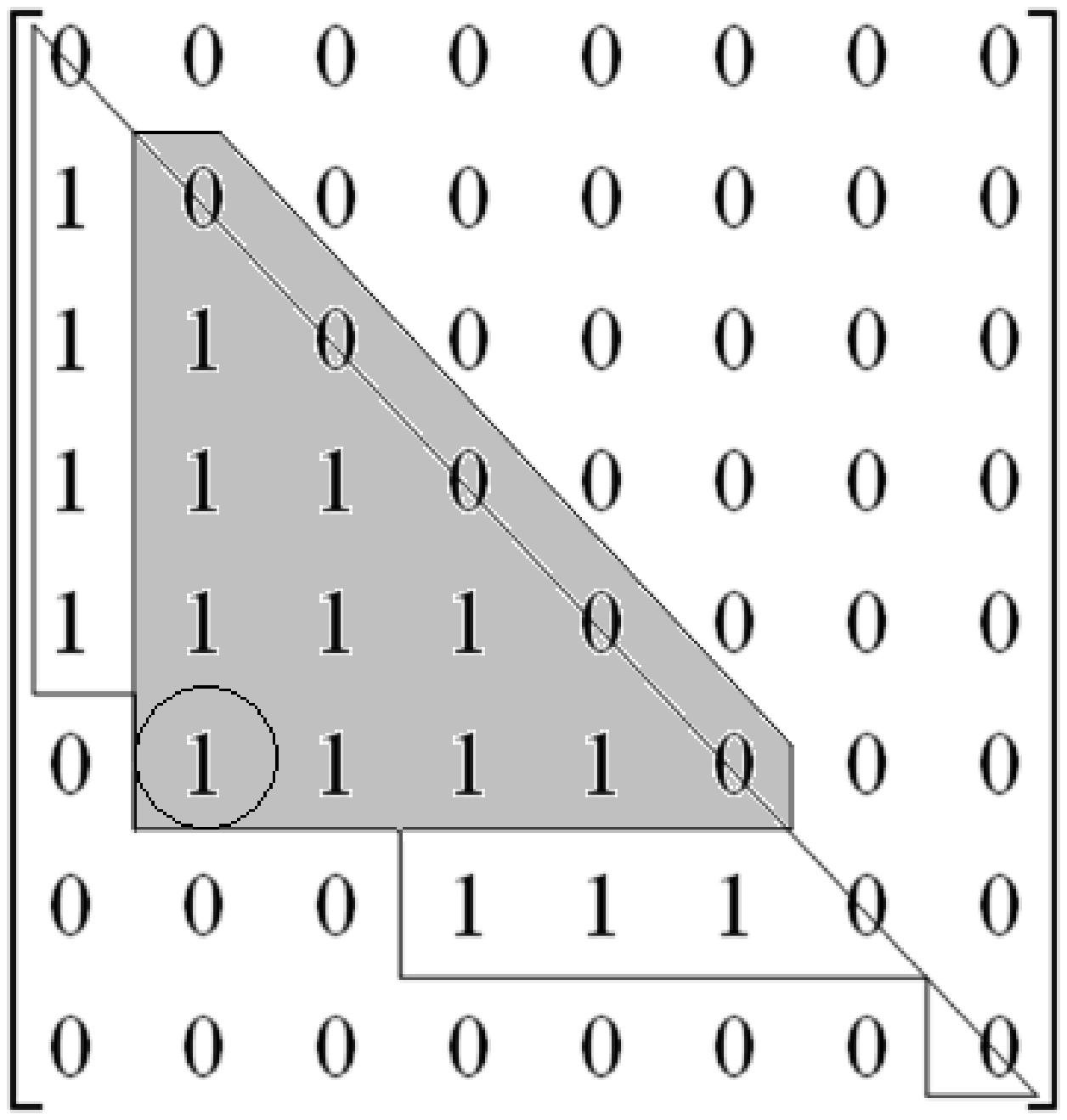}}}
\caption{(a) The NIR matrix $H_G $ of an interval graph $G$, (b) The SNIR
matrix $H_{G^{\prime }} $ of a proper interval graph $G^{\prime }$.}
\label{Fig_1}
\end{figure}

Recall that the left and the right endpoints of the $i^{th}$ interval in the
SNIR form of $G$ correspond to the $i^{th}$ and the $\left( {x_i +i}
\right)^{th}$ elements of the $i^{th}$ column of $H_G $ respectively.
Therefore, due to Definition \ref{Def_3}, it holds that $x_i +i\ge x_j +j$
for $i>j$. Consequently, any stair of $H_G$ consists of unit matrix
elements, except of the diagonal elements of $H_G$, while the corresponding
pick is the lower most left matrix entry of this stair. As it is seen in
Figure~\ref{Fig_1b}, the SNIR matrix $H_G$ has a stair-shape and equals the
union of all its stairs. A stair of $H_G$ can be also recognized in this
figure, where the corresponding pick is marked with a circle. 

\begin{lem}
\label{Lem_5} An arbitrary graph is a proper interval graph iff it can be
represented by the SNIR form.
\end{lem}

\begin{proof}
Due to Lemma~\ref{Lem_4}, any proper interval graph can be 
represented by the SNIR form. Conversely, the SNIR form is clearly a set of 
intervals, where no one of which includes strictly another one, 
i.e., it is a realization of a proper interval graph.
\end{proof}

\begin{lem}
\label{Lem_6} Any stair of the SNIR matrix $H_G $ corresponds bijectively to
a maximal clique in $G$.
\end{lem}

\begin{proof}
Due to Lemma~\ref{Lem_3}, every maximal clique of $G$ corresponds bijectively 
to a row of $H_G$, in which at least one of its unit elements or its zero diagonal element does not 
have any chain of $1$'s below it. However, since $G$ is a proper interval graph and due to 
Definition~\ref{Def_5}, such a row corresponds bijectively to a pick of $H_G$ and therefore to 
a stair of it, as it is shown in Figure~\ref{Fig_1b}. 
\end{proof}

\section{The k-cluster problem on proper interval graphs}

\label{Cluster_Pr} Due to Lemma~\ref{Lem_5}, a proper interval graph $G$ is
equivalent to an SNIR matrix $H_{G}$. Denote by $S_{1},S_{2},...,S_{m}$, $%
m\leq n-1$, the stairs of $H_{G}$, numbered from the top to the bottom. Due
to Lemma~\ref{Lem_6} these stairs correspond bijectively to the maximal
cliques $Q_{1},Q_{2},...,Q_{m}$, of $G$. Denote for simplicity $%
S_{0}:=\emptyset $ and $Q_{0}:=\emptyset $. Every stair $S_{i}$ constitutes
together with its previous stairs $S_{1},S_{2},...,S_{i-1}$ a submatrix $%
H_{i}:=H_{G_{i}}$ of $H_{G}$ that is equivalent to the subgraph $%
G_{i}:=\bigcup\nolimits_{\ell =1}^{i}{Q_{\ell }}$ of $G$, which remains also
a proper interval graph. In particular, $H_{m}=H_{G}$ is equivalent to $%
G_{m}=G$. We develop further a dynamic programming algorithm for the
j-cluster problem on $G_{i}$, which makes use of the optimal solutions of
the q-cluster problems on $G_{i-1}$, for $q=1,2,...,j$. The critical
observation here is that the arbitrary $i^{th}$ stair $S_{i}$ of $H_{G}$
contains at least one row that does not belong to the previous stair $%
S_{i-1} $, i.e. $S_{i}\setminus S_{i-1}\neq \emptyset $ and therefore $%
Q_{i}\setminus Q_{i-1}\neq \emptyset $. Suppose that the pick of $S_{i}$ is
the matrix element $H_{G}\left( {a_{i},b_{i}}\right) $. Then, the maximal
clique $Q_{i}$ has $\left\vert {Q_{i}}\right\vert =a_{i}-b_{i}+1$ nodes,
namely the $b_{i}^{th},\left( {b_{i}+1}\right) ^{th},...,a_{i}^{th}$ ones.

Denote now by $f_{i}\left( {j,x,x}^{\prime }\right) $ the value of an
optimal solution of the j-cluster problem on $G_{i}$, including \emph{exactly%
} $x$ nodes of the clique $Q_{i}\setminus Q_{i-1}$ and \emph{exactly} $%
x^{\prime }$ nodes of the clique $Q_{i}\cap Q_{i-1}$. Clearly, $0\leq x\leq {%
\left\vert {Q_{i}\setminus Q_{i-1}}\right\vert }$, $0\leq x^{\prime }\leq {%
\left\vert {Q_{i}\cap Q_{i-1}}\right\vert }$ and $x+x^{\prime }\leq j$.
Then, the value of an optimal solution of the j-cluster problem on $G_{i}$
is $f_{i}\left( j\right) =\mathop{\max }\limits_{x,x^{\prime }}\left\{ {%
f_{i}\left( {j,x,x}^{\prime }\right) }\right\} $. Note that obviously for
the j-cluster problem on a single stair $H_{1}=S_{1}$ we should require that 
$x^{\prime }=0$ and $x=j$, as also that $Q_{1}$ has at least $j$ nodes,
since otherwise we should include also $j-x>0$ nodes of $Q_{0}=\emptyset $,
which is a contradiction. Therefore, the following initial conditions hold
for $i=1$ and $j=1,2,...,k$: 
\begin{equation}
f_{1}\left( {j,x,0}\right) =%
\begin{cases}
\left( {{\begin{array}{*{20}c} j \hfill \\ 2 \hfill \\ \end{array}}}\right) ,
& \text{if }x=j\leq \left\vert {Q_{1}}\right\vert \\ 
-\infty , & \text{otherwise}%
\end{cases}
\label{eq9}
\end{equation}

\begin{figure}[h]
\centering
\mbox{
      \subfigure[]{        
        \label{Fig_2a}
        \includegraphics[height=7.5cm]{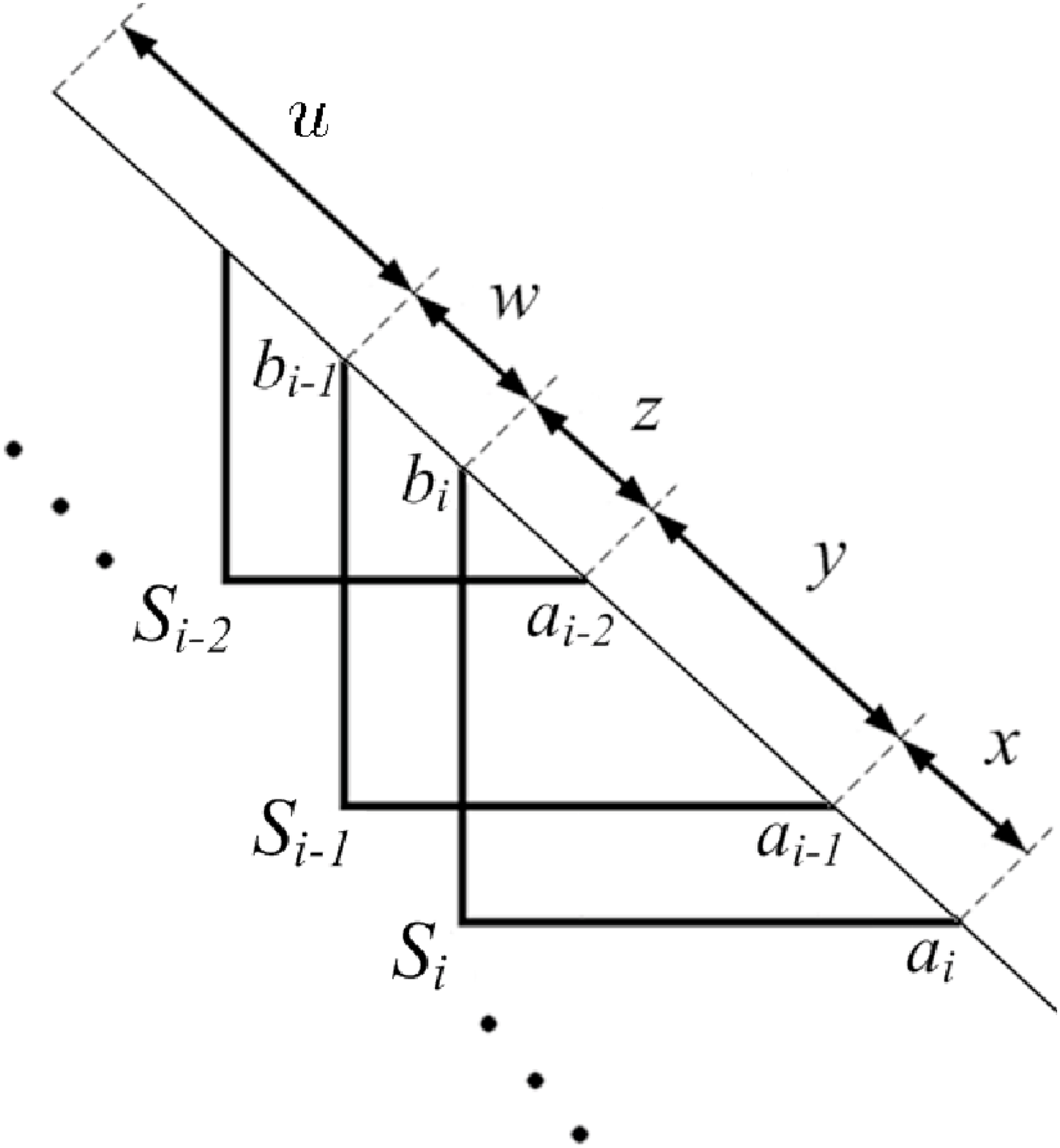}}
        
      \subfigure[]{
        \label{Fig_2b}
        \includegraphics[height=7.5cm]{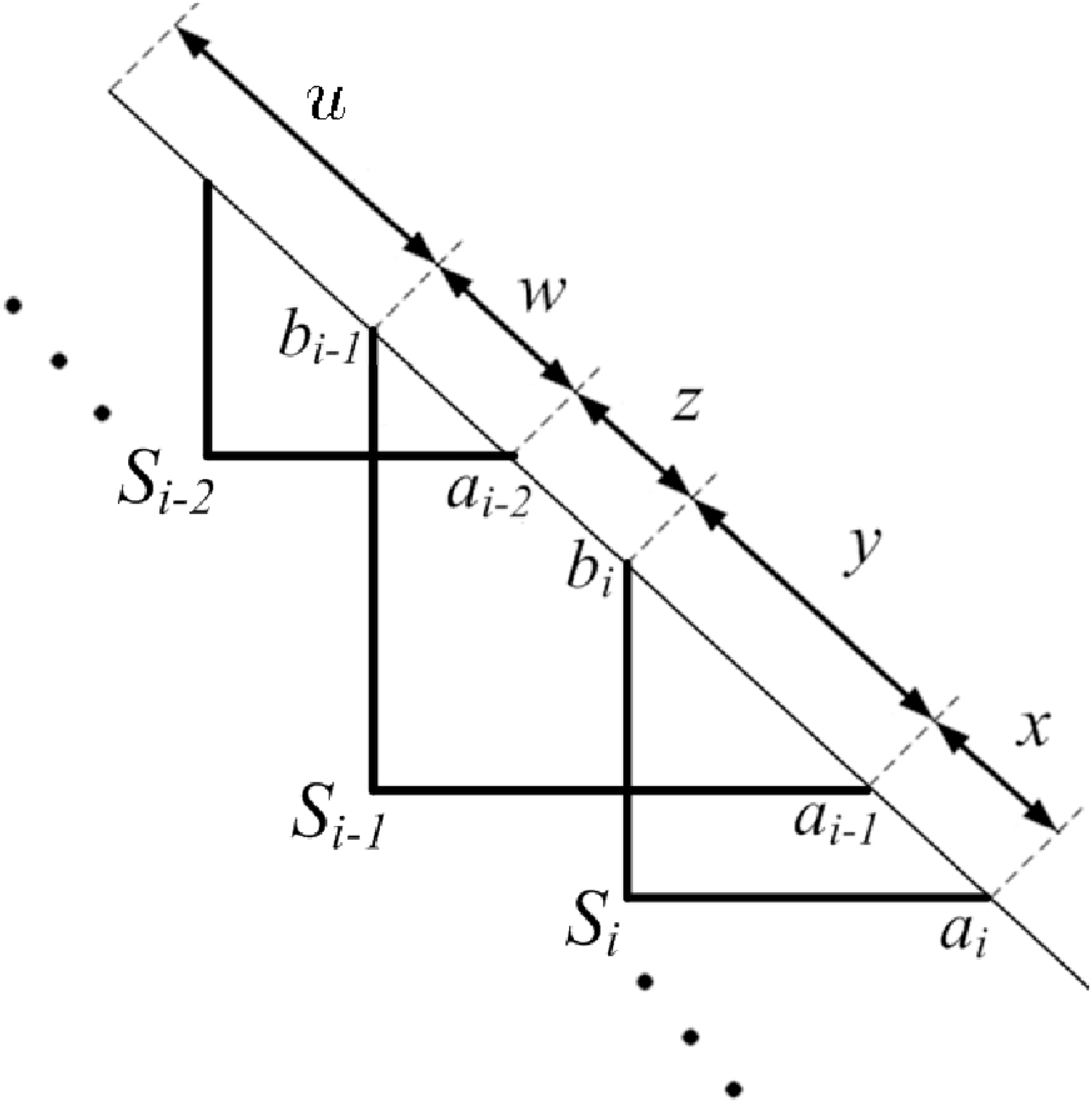}}}
\caption{The split of the SNIR matrix $H_{G}$ for the recursion of the
k-cluster problem on a proper interval graph $G$, in the cases (a) $%
S_{i}\cap S_{i-2}\neq \emptyset $ and (b) $S_{i}\cap S_{i-2}=\emptyset $.}
\label{Fig_2}
\end{figure}

If $j\leq \left\vert {Q_{i}}\right\vert $, then any subclique of $Q_{i}$ on $%
j$ nodes is clearly an optimal solution. Otherwise, consider the case $%
j>\left\vert {Q_{i}}\right\vert $. The recursive computation of $f_{i}\left( 
{j,x,x}^{\prime }\right) $, which is presented below, makes use of the
values $f_{i-1}\left( {q,r,r}^{\prime }\right) $ for $q=1,2,...,j$, where $%
x=\left\vert Q_{i}\setminus Q_{i-1}\right\vert $, $x^{\prime }=\left\vert
Q_{i}\cap Q_{i-1}\right\vert $, $r=\left\vert Q_{i-1}\setminus
Q_{i-2}\right\vert $ and $r^{\prime }=\left\vert Q_{i-1}\cap
Q_{i-2}\right\vert $. We distinguish the cases $Q_{i}\cap Q_{i-2}\neq
\emptyset $ and $Q_{i}\cap Q_{i-2}=\emptyset $, or equivalently $S_{i}\cap
S_{i-2}\neq \emptyset $ and $S_{i}\cap S_{i-2}=\emptyset $. In the case $%
Q_{i}\cap Q_{i-2}\neq \emptyset $ an optimal solution may include $y$ nodes
of $Q_{i-1}\setminus Q_{i-2}$, $z$ nodes of $Q_{i}\cap Q_{i-2}$, $w$ nodes\
of $Q_{i-1}\setminus Q_{i}$ and $u$ nodes of the remaining part of $G_{i}$.
In the opposite case $Q_{i}\cap Q_{i-2}=\emptyset $, an optimal solution may
include $y$ nodes of $Q_{i}\cap Q_{i-1}$, $z$ nodes of $Q_{i-1}\setminus
\left( {Q_{i}\cup Q_{i-2}}\right) $, $w$ nodes of $Q_{i-1}\cap Q_{i-2}$ and $%
u$ nodes of the remaining part of $G_{i}$. Both situations are illustrated
in Figure~\ref{Fig_2}. As it can be easily verified, for all these sets the
following hold: 
\begin{equation}
\begin{array}{l}
\text{Case }Q_{i}\cap Q_{i-2}\neq \emptyset : \\*[5pt] 
0\leq x\leq x_{0}:=\left\vert {Q_{i}\setminus Q_{i-1}}\right\vert \\* 
\quad \quad \,\,\,\,=a_{i}-a_{i-1} \\* 
0\leq y\leq y_{1}:=\left\vert {Q_{i-1}\setminus Q_{i-2}}\right\vert \\* 
\quad \quad \,\,\,\,=a_{i-1}-a_{i-2} \\* 
0\leq z\leq z_{1}:=\left\vert {Q_{i}\cap Q_{i-2}}\right\vert \\* 
\quad \quad \,\,\,\,=a_{i-2}-b_{i}+1 \\* 
0\leq w\leq w_{1}:=\left\vert {Q_{i-1}\setminus Q}_{i}\right\vert \\* 
\quad \quad \,\,\,\,=b_{i}-b_{i-1} \\* 
0\leq u\leq u_{1}:=b_{i-1}-1 \\ 
\end{array}%
\quad 
\begin{array}{l}
\text{Case }Q_{i}\cap Q_{i-2}=\emptyset : \\*[5pt] 
0\leq x\leq x_{0}=\left\vert {Q_{i}\setminus Q_{i-1}}\right\vert \\* 
\quad \quad \,\,\,\,=a_{i}-a_{i-1} \\* 
0\leq y\leq y_{2}:=\left\vert {Q_{i}\cap Q_{i-1}}\right\vert \\* 
\quad \quad \,\,\,\,=a_{i-1}-b_{i}+1 \\* 
0\leq z\leq z_{2}:=\left\vert {Q_{i-1}\setminus \left( {Q_{i}\cup Q_{i-2}}%
\right) }\right\vert \\* 
\quad \quad \,\,\,\,=b_{i}-a_{i-2}-1 \\* 
0\leq w\leq w_{2}:=|{Q_{i-1}\cap Q_{i-2}|} \\* 
\quad \quad \,\,\,\,=a_{i-2}-b_{i-1}+1 \\* 
0\leq u\leq u_{2}:=b_{i-1}-1 \\ 
\end{array}
\label{eq6}
\end{equation}%
The case $Q_{i}\cap Q_{i-2}\neq \emptyset $ occurs exactly when $b_{i}\leq
a_{i-2}$, i.e. $H\left( {a_{i-2}-b_{i}}\right) =1$, while the opposite case $%
Q_{i}\cap Q_{i-2}=\emptyset $ occurs exactly when $H\left( {b_{i}-a_{i-2}-1}%
\right) =1$. Thus, since $x,y,z,w$ and $u$ add up to $j$, we can summarize
the relations in (\ref{eq6}) to the following, for the general case: 
\begin{align}
& 0\leq x\leq x_{0}  \notag \\
& 0\leq y\leq y_{1}\cdot H\left( {a_{i-2}-b_{i}}\right) +{y}_{2}\cdot
H\left( {b_{i}-a_{i-2}-1}\right)  \notag \\
& 0\leq z\leq {z}_{1}\cdot H\left( {a_{i-2}-b_{i}}\right) +{z}_{2}\cdot
H\left( {b_{i}-a_{i-2}-1}\right)  \label{eq7} \\
& 0\leq w\leq w_{1}\cdot H\left( {a_{i-2}-b_{i}}\right) +w_{2}\cdot H\left( {%
b_{i}-a_{i-2}-1}\right)  \notag \\
& 0\leq u\leq u_{1}\cdot H\left( {a_{i-2}-b_{i}}\right) +u_{2}\cdot H\left( {%
b_{i}-a_{i-2}-1}\right)  \notag \\
& x+y+z+w+u=j  \notag
\end{align}%
For simplicity, let $\zeta _{1}=z\cdot {H(a}_{i-2}-b_{i}{)}$ and $\zeta
_{2}=z\cdot {H\left( \ {b_{i}-a_{i-2}-1}\right) \ }$. Now, the value $%
f_{i}\left( {j,x,x}^{\prime }\right) $ can be computed by using the top-down
approach of the following equation, for both cases $Q_{i}\cap Q_{i-2}\neq
\emptyset $ and $Q_{i}\cap Q_{i-2}=\emptyset $:

\begin{equation}
f_{i}\left( {j,x,y+}\zeta _{1}\right) =%
\begin{cases}
\left( {{\begin{array}{*{20}c} j \hfill \\ 2 \hfill \\ \end{array}}}\right) ,
& \text{if }{x+y+}\zeta _{1}=\text{$j\leq \left\vert {Q_{i}}\right\vert $ }
\\ 
\begin{split}
\mathop {\max }\limits_{y,z,w,u\in (\ref{eq7})}{\bigg\{}& f_{i-1}\left( {%
j-x,y+}\zeta _{2},\zeta _{1}+w\right) \\
& \quad +\left( {{\begin{array}{*{20}c} x \hfill \\ 2 \hfill \\ \end{array}}}%
\right) +x\left( {y+}\zeta _{1}\right) {\bigg \}},
\end{split}
& \text{otherwise}%
\end{cases}
\label{eq8}
\end{equation}%
Finally, the dynamic programming Algorithm 1 returns the value of an optimal
solution of the k-cluster problem on $G$. After applying some necessary
modifications, it will return the optimal solution, instead of its value.

\begin{tabbing}
\textbf{Algorithm} Proper-Interval-k-cluster problem($G$):\\*
\textbf{\emph{Input:}} An arbitrary realization of a proper interval graph $G$\\*
\textbf{\emph{Output:}} The value of an optimal solution of the k-cluster problem on $G$\\*[-15pt]
\end{tabbing}

\begin{enumerate}
\item \vspace{-0.3cm} Construct the SNIR matrix $H_G$. Let that $H_G$ has
the $m$ stairs $S_1 ,S_2 ,...,S_m $ that correspond to the maximal cliques $%
Q_1 ,Q_2 ,...,Q_m $ of $G$

\item \vspace{-0.3cm} \textbf{If} $m=1$ \textbf{Then Return} $%
f_{1}(k)=f_{1}\left( {k,k,0}\right) $, computed from (\ref{eq9});\\*[0pt]
\textbf{Else Return} $f_{m}(k)=\max \{{f_{m}\left( {k,x,x}^{\prime }\right)
:0\leq x\leq {\left\vert {Q_{i}\setminus Q_{i-1}}\right\vert ,0\leq x}}%
^{\prime }{{\leq {\left\vert {Q_{i}\cap Q_{i-1}}\right\vert ,x+x}}}^{\prime
}\leq k{\}}$, computed from (\ref{eq8})\\*[-20pt]
\end{enumerate}

\vspace{-0.2cm} Algorithm 1: The value of an optimal solution of the
k-cluster problem on the proper interval graph $G$.

\begin{thm}
\label{Thm_2} The k-cluster problem is solvable in $O\left( {nk^{5}}\right) $
time on proper interval graphs.
\end{thm}

\begin{proof}
The computation of a single $f_{i}\left( j\right) $ in the Algorithm 1 takes
at most $O\left( {j^{4}}\right) =O\left( {k^{4}}\right) $ time due to the
combinations of the $x,y,z,w,u$, such that they sum up to $j$, since $x,y,z$
and $w$ may vary and $u=j-x-y-z-w$ is then uniquely determined by them.
Every $f_{i}\left( j\right) $ is computed for all $i\in \{1,2,...,m\}$ and $%
j\in \{1,2,...,k\}$, i.e., altogether at most $m\cdot k=O\left( {nk}\right) $
quantities are computed. Thus, since any proper interval graph can be
recognized and converted to the SNIR form in linear time, the k-cluster
problem can be solved in $O\left( {nk^{5}}\right) $ time on any proper
interval graph.
\end{proof}Note that in the presented analysis the subgraph that corresponds
to the obtained optimal solution is not necessarily connected. Lemma \ref%
{Lem_7} proposes a modification to the Algorithm 1, in order to find an
optimal solution, under the additional constraint of connectivity.

\begin{lem}
\label{Lem_7} The Algorithm 1 returns the value of an optimal solution of
the k-cluster problem on proper interval graphs, under the additional
constraint of connectivity, if the following additional condition to (\ref%
{eq7}) is required: 
\begin{equation}
y+\zeta _{1}\geq 1,\text{ if }x>0\text{.}  \label{eq10}
\end{equation}%
After this modification, the runtime of the proposed algorithm remains $%
O\left( {nk^{5}}\right) $.
\end{lem}

\begin{proof}
The proof is done by induction. If $i=1$, then the obtained solution is
always connected, as an induced subgraph of a clique. Suppose now that $i>1$
and $x>0$. It follows that we use $x\geq 1$ nodes of $Q_{i}$, which are not
included in $Q_{j}$, for any $j<i$. Therefore, in order to construct a
connected subgraph, it is equivalent to require that at least one node of 
$Q_{i}\cap G_{i-1}=Q_{i}\cap Q_{i-1}$ is included, i.e., a node which is
simultaneously connected to the $x$ nodes of $Q_{i}\setminus Q_{i-1}$ and to
at least one node of the remaining graph $G_{i-1}$. However, as described
above, we include in the constructed subgraph exactly $y+z$ nodes of 
$Q_{i}\cap Q_{i-1}$ if $Q_{i}\cap Q_{i-2}\neq \emptyset $ and exactly $y$
nodes of $Q_{i}\cap Q_{i-1}$ if $Q_{i}\cap Q_{i-2}=\emptyset $. Namely, we
include exactly $y+\zeta _{1}$ nodes of $Q_{i}\cap Q_{i-1}$ in the general
case. Therefore, in order to construct a connected subgraph, it is
equivalent to require that $y+\zeta _{1}\geq 1$. Finally, the asymptotic
complexity of the proposed algorithm remains obviously unchanged, when
requiring the additional condition (\ref{eq10}) to the conditions (\ref{eq7}).
\end{proof}

\section{The k-cluster problem on interval graphs}

\label{Cluster_Int} In this section we propose a polynomial dynamic
programming algorithm for the k-cluster problem on interval graphs, whose
complexity status was an open question \cite{Corneil84}. The proposed
algorithm constitutes a generalization of Algorithm 1 for proper interval
graphs. Due to Lemma~\ref{Lem_2}, an interval graph $G$ is equivalent to a
NIR matrix $H_{G}$. In the following consider an interval graph $G$ on $n$
nodes, as well as its NIR matrix $H_{G}$.

Due to Lemma~\ref{Lem_3} any maximal clique of $G$ corresponds bijectively
to a row of the NIR matrix $H_{G}$, in which at least one of its unit
elements or its zero diagonal element does not have any chain of $1$'s below
it. The maximal clique, which refers to such a row, contains all intervals,
i.e. nodes, which correspond to the unit elements and the zero diagonal
element of this row. Denote these maximal cliques of $G$ by $%
Q_{1},Q_{2},...,Q_{m}$, $m\leq n-1$, numbered from the top to the bottom, as
well as $Q_{0}:=\emptyset $. Suppose also that the maximal clique $Q_{\ell }$
occurs at the $a_{\ell }^{th}$ row of $H_{G}$ and denote by $\left\vert {%
Q_{\ell }}\right\vert $ the number of nodes of $Q_{\ell }$. It holds clearly
that $Q_{i}\setminus Q_{i-1}\neq \emptyset $ for all $i=1,2,...,m$. Every
maximal clique $Q_{i}$ constitutes together with its previous maximal
cliques $Q_{1},Q_{2},...,Q_{i-1}$ a subgraph $G_{i}$ of $G$, which remains
also an interval graph. Similarly to Section \ref{Cluster_Pr} for the proper
interval graphs, we develop further a dynamic programming algorithm for the
j-cluster problem on $G_{i}$, which makes use of the optimal solutions of
the q-cluster problems on $G_{i-1}$, for $q=1,2,...,j$.

An optimal solution may include $y$ nodes of $({Q_{i}\cap Q_{i-1})\setminus
Q_{i-2}}$, $z$ nodes of ${Q_{i-1}\setminus (Q}_{i}\cup {Q_{i-2})}$, $w$
nodes\ of ${Q_{i}\cap Q}_{i-2}$, $u$ nodes of $({Q_{i-1}\cap Q}%
_{i-2})\setminus Q_{i}$ and $v$ nodes of the remaining part of $G_{i}$, as
it is illustrated in Figure~\ref{Fig_3}. We compute in Appendix \ref%
{Split_Int} the split of the NIR matrix $H_{G}$ and we obtain the following
relations for the variables $x,y,z,w,u$ and $v$:

\begin{equation}
\begin{array}{l}
0\leq x\leq |{Q_{i}\setminus Q_{i-1}|}=a_{i}-a_{i-1} \\* 
0\leq y\leq |({Q_{i}\cap Q_{i-1})\setminus Q_{i-2}|} \\* 
\quad \quad \,\,\,\,=\sum\nolimits_{\ell =a_{i-2}+1}^{a_{i-1}}{H\left( {\ell
+x_{\ell }-a_{i}}\right) } \\* 
0\leq z\leq |{Q_{i-1}\setminus (Q}_{i}\cup {Q_{i-2})|} \\* 
\quad \quad \,\,\,\,=a_{i-1}-a_{i-2}-\sum\nolimits_{\ell
=a_{i-2}+1}^{a_{i-1}}{H\left( {\ell +x_{\ell }-a_{i}}\right) } \\* 
0\leq w\leq |{Q_{i}\cap Q}_{i-2}|=\sum\nolimits_{\ell =1}^{a_{i-2}}{H\left( {%
\ell +x_{\ell }-a_{i}}\right) } \\ 
0\leq u\leq |({Q_{i-1}\cap Q}_{i-2})\setminus Q_{i}| \\* 
\quad \quad \,\,\,\,=\sum\nolimits_{\ell =1}^{a_{i-2}}{H\left( {\ell
+x_{\ell }-a_{i-1}}\right) \cdot H\left( a_{i}-{\ell -x_{\ell }-1}\right) }
\\* 
0\leq v\leq a_{i-2}-\sum\nolimits_{\ell =1}^{a_{i-2}}{H\left( {\ell +x_{\ell
}-a_{i}}\right) -} \\ 
\,\,\,\,\,-\sum\nolimits_{\ell =1}^{a_{i-2}}{H\left( {\ell +x_{\ell }-a_{i-1}%
}\right) \cdot H\left( a_{i}-{\ell -x_{\ell }-1}\right) } \\ 
x+y+z+w+u+v=j \\ 
\end{array}
\label{eq11}
\end{equation}%
Now, the value $f_{i}\left( {j,x,x}^{\prime }\right) $ can be computed by
using the top-down approach of the following equation:

\begin{equation}
f_{i}\left( {j,x,y+}w\right) =%
\begin{cases}
\left( {{\begin{array}{*{20}c} j \hfill \\ 2 \hfill \\ \end{array}}}\right) ,
& \text{if }{x+y+}w=\text{$j\leq \left\vert {Q_{i}}\right\vert $ } \\ 
\begin{split}
\mathop {\max }\limits_{y,z,w,u,v\in (\ref{eq11})}{\bigg\{}& f_{i-1}\left( {%
j-x,y+}z,w+u\right) \\
& \quad +\left( {{\begin{array}{*{20}c} x \hfill \\ 2 \hfill \\ \end{array}}}%
\right) +x\left( {y+}w\right) {\bigg \}},
\end{split}
& \text{otherwise}%
\end{cases}
\label{eq12}
\end{equation}%
Finally, the dynamic programming Algorithm 2, similarly to Algorithm 1,
returns the value of an optimal solution of the k-cluster problem on $G$.
After applying some necessary modifications, it will return the optimal
solution, instead of its value. 
\begin{tabbing}
\textbf{Algorithm} Interval-k-cluster problem($G$):\\*
\textbf{\emph{Input:}} An arbitrary realization of an interval graph $G$\\*
\textbf{\emph{Output:}} The value of an optimal solution of the k-cluster problem on $G$\\*[-15pt]
\end{tabbing}

\begin{enumerate}
\item \vspace{-0.3cm} Construct the NIR matrix $H_G$. Let that $G$ has the $%
m $ maximal cliques $Q_1 ,Q_2 ,...,Q_m $

\item \vspace{-0.3cm} \textbf{If} $m=1$ \textbf{Then Return} $%
f_{1}(k)=f_{1}\left( {k,k,0}\right) $, computed from (\ref{eq9});\\*[0pt]
\textbf{Else Return} $f_{m}(k)=\max \{{f_{m}\left( {k,x,x}^{\prime }\right)
:0\leq x\leq {\left\vert {Q_{i}\setminus Q_{i-1}}\right\vert ,0\leq x}}%
^{\prime }{{\leq {\left\vert {Q_{i}\cap Q_{i-1}}\right\vert ,x+x}}}^{\prime
}\leq k{\}}$, computed from (\ref{eq12})\\*[-20pt]
\end{enumerate}

\vspace{-0.2cm} Algorithm 2: The value of an optimal solution of the
k-cluster problem on the interval graph $G$. 
\begin{figure}[h]
\centering
\includegraphics[height=7.3cm]{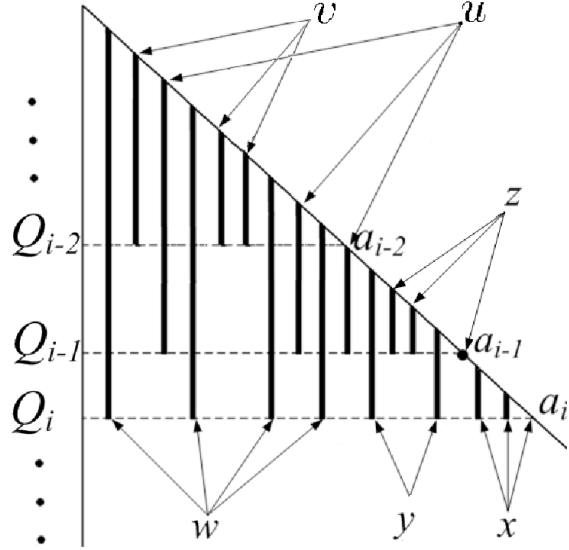}
\caption{The split of the NIR matrix $H_G $ for the recursion of the
k-cluster problem on an interval graph $G$.}
\label{Fig_3}
\end{figure}

\begin{thm}
\label{Thm_3} The k-cluster problem is solvable in $O\left( {nk^{6}}\right) $
time on interval graphs.
\end{thm}

\begin{proof}
The computation of a single $f_{i}\left( j\right) $ in the Algorithm 2 takes
at most $O\left( {j^{5}}\right) =O\left( {k^{5}}\right) $ time due to the
combinations of the $x,y,z,w,u,v$, such that they sum up to $j$, since $x,y,z,w$
and $u$ may vary and $v=j-x-y-z-w-u$ is then uniquely determined by them.
Every $f_{i}\left( j\right) $ is computed for all $i\in \{1,2,...,m\}$ and $j\in
\{1,2,...,k\}$, i.e., altogether at most $m\cdot k=O\left( {nk}\right) $
quantities are computed. Thus, since any interval graph can be recognized
and converted to the NIR form in linear time, the k-cluster problem can be
solved in $O\left( {nk^{6}}\right) $ time on any interval graph.
\end{proof}

\begin{lem}
\label{Lem_8} The proposed algorithm returns the value of an optimal
solution of the k-cluster problem on interval graphs, under the additional
constraint of connectivity, if the following additional condition is
required to the conditions (\ref{eq12}): 
\begin{equation}
y+w\geq 1,\text{ if }x>0\text{.}  \label{eq14}
\end{equation}%
After this modification, the runtime of the proposed algorithm remains $%
O\left( {nk^{6}}\right) $.
\end{lem}

\begin{proof}
The proof is done by induction. If $i=1$, then the obtained solution is
always connected, as an induced subgraph of a clique. Suppose now that $i>1$
and $x>0$. It follows that we use $x\geq 1$ nodes of $Q_{i}$, which are not
included in $Q_{j}$, for any $j<i$. Therefore, in order to construct a
connected subgraph, it is equivalent to require that at least one node of 
$Q_{i}\cap G_{i-1}=Q_{i}\cap Q_{i-1}$ is included, i.e., a node which is
simultaneously connected to the $x$ nodes of $Q_{i}\setminus Q_{i-1}$ and to
at least one node of the remaining graph $G_{i-1}$. However, as described
above, we include in the constructed subgraph exactly $y+w$ nodes of 
$Q_{i}\cap Q_{i-1}$. Therefore, in order to construct a connected subgraph,
it is equivalent to require that $y+w\geq 1$. Finally, the asymptotic
complexity of the proposed algorithm remains obviously unchanged, when
requiring the additional condition (\ref{eq14}) to the conditions (\ref{eq11}).
\end{proof}

\section{Conclusions}

\label{Concl} In this paper an efficient matrix representation that
characterizes the interval graphs, as well as its restriction on the proper
interval graphs is used, which leads to a simple polynomial time algorithm
for the k-cluster problem on these classes of graphs. This problem is known
to be NP-hard on an arbitrary graph, as a generalization of the maximum
clique problem, as well as on the chordal graphs. In contrary, its
complexity on interval and proper interval graphs was an open question.

\subsubsection*{Acknowledgment}
I wish to thank Professor Philippe Baptiste and Dr.~Maxim Sviridenko 
for reading the manuscript and improving the presentation.

\bibliographystyle{unsrt}
\bibliography{References}

\begin{thebibliography}{10}

\bibitem{Hsu93}
W.L. Hsu.
\newblock A simple test for interval graphs.
\newblock In {\em WG '92: Proceedings of the 18th International Workshop on
  Graph-Theoretic Concepts in Computer Science}, pages 11--16, London, 1993.
  Springer-Verlag.

\bibitem{Corneil98}
D.~G. Corneil, S.~Olariu, and L.~Stewart.
\newblock The ultimate interval graph recognition algorithm?
\newblock In {\em SODA '98: Proceedings of the ninth annual ACM-SIAM symposium
  on Discrete algorithms}, pages 175--180, Philadelphia, PA, USA, 1998. Society
  for Industrial and Applied Mathematics.

\bibitem{Corneil95}
D.~Corneil, H.~Kim, S.~Natarajan, S.~Olariu, and A.P. Sprague.
\newblock Simple linear time recognition of unit interval graphs.
\newblock {\em Inform. Process. Lett.}, 55:99--104, 1995.

\bibitem{Golumbic04}
M.C. Golumbic and A.N. Trenk.
\newblock {\em Tolerance graphs}.
\newblock Cambridge University Press, Cambridge, 2004.

\bibitem{Carrano88}
A.V. Carrano.
\newblock Establishing the order to human chromosome-specific {DNA} fragments.
\newblock In A.~D. Woodhead and B.~J. Barnhart, editors, {\em Biotechnology and
  the Human Genome}, pages 37--50. Plenum Press, New York, 1988.

\bibitem{Gupta82}
U.I. Gupta, D.T. Lee, and J.Y.T. Leung.
\newblock Efficient algorithms for interval graphs and circular-arc graphs.
\newblock {\em Networks}, pages 459--467, 1982.

\bibitem{Ju92}
Ju~Yuan Hsiao and Chuan~Yi Tang.
\newblock An efficient algorithm for finding a maximum weight 2-independent set
  on interval graphs.
\newblock {\em Inf. Process. Lett.}, 43(5):229--235, 1992.

\bibitem{Chang93}
M.S. Chang, S.L. Peng, and J.L. Liaw.
\newblock Deferred-query - an efficient approach for problems on interval and
  circular-arc graphs (extended abstract).
\newblock In {\em WADS}, pages 222--233, 1993.

\bibitem{Corneil84}
D.G. Corneil and Y.~Perl.
\newblock Clustering and domination in perfect graphs.
\newblock {\em Discrete Applied Mathematics}, 9:27--39, 1984.

\bibitem{Holzapfel03}
K.~Holzapfel, S.~Kosub, M.G. Maa{\ss}, and H.~T{"a}ubig.
\newblock The complexity of detecting fixed-density clusters.
\newblock In {\em Proceedings of the 5th Italian Conference on Algorithms and
  Complexity (CIAC'2003)}, volume 2653, pages 201--212, Berlin, 2003.
  Springer-Verlag.
\newblock Lecture Notes in Computer Science.

\bibitem{Liazi05}
M.~Liazi, I.~Milis, and V.~Zissimopoulos.
\newblock On the complexity of the densest/heaviest k-subgraph problem.
\newblock Preprint submitted to Elsevier Science, June 2005.

\bibitem{Mertzios07}
G.B. Mertzios.
\newblock A matrix characterization of interval and proper interval graphs.
\newblock {\em Applied Mathematics Letters}, 2007.
\newblock To appear.

\end{thebibliography}

\newpage 
\begin{appendix}
\section{The split of the NIR matrix $H_G$}
\label{Split_Int}
We remind at first that it is assumed that the maximal clique $Q_{\ell }$
occurs at the $\ell ^{th}$ row of $H_{G}$, for $\ell =1,2,...,m$. Suppose
that $a_{i-1}<\ell \leq a_{i}$. If the chain of $1$'s under the $\ell ^{th}$
diagonal element of $H_{G}$ stops at a row, which is higher than the $%
a_{i}^{th}$ one, then a maximal clique would occur between $Q_{i-1}$ and $%
Q_{i}$, which is a contradiction. Thus, the chain under the $\ell ^{th}$
diagonal element stops either at the $a_{i}^{th}$ row, or even lower.
Suppose now that $\ell \leq a_{i-1}$. If $\ell \in Q_{i}$, then also $\ell
\in Q_{i-1}$, since the chain under the $\ell ^{th}$ diagonal element stops
either at the $a_{i}^{th}$ row, or even lower, i.e. strictly lower than the $%
a_{i-1}^{th}$ row. Therefore, the elements of $Q_{i}\setminus Q_{i-1}$ are
exactly the $\left( {a_{i-1}+1}\right) ^{th},...,a_{i}^{th}$ diagonal
elements. Thus, 
\begin{equation}
\left\vert {Q_{i}\setminus Q_{i-1}}\right\vert =a_{i}-a_{i-1}  \label{eqA1}
\end{equation}%
In order to compute the value $|({Q_{i}\cap Q_{i-1})\setminus Q_{i-2}|}$, we
have to compute how many of the $1^{st},2^{nd},...,a_{i-1}^{th}$ diagonal
elements belong to $Q_{i}$ and to $Q_{i-1}$, but not to $Q_{i-2}$. For $%
1\leq \ell \leq a_{i-1}$, the arbitrary $\ell ^{th}$ diagonal element
belongs to $Q_{i}$ exactly when its chain of $1$'s reaches the $a_{i}^{th}$
row, i.e. exactly when $\ell +x_{\ell }\geq a_{i}$, or equivalently $H\left( 
{\ell +x_{\ell }-a_{i}}\right) =1$. In this case, it belongs also to $Q_{i-1}
$, since $a_{i-1}<a_{i}$. Further, for $1\leq \ell \leq a_{i-2}$, if $%
H\left( {\ell +x_{\ell }-a_{i}}\right) =1$ then the $\ell ^{th}$ diagonal
element belongs also to $Q_{i-2}$ and therefore not to $({Q_{i}\cap
Q_{i-1})\setminus Q_{i-2}}$. It follows that

\begin{equation}
|({Q_{i}\cap Q_{i-1})\setminus Q_{i-2}|}=\sum\nolimits_{\ell
=a_{i-2}+1}^{a_{i-1}}{H\left( {\ell +x_{\ell }-a_{i}}\right) }  \label{eqA2}
\end{equation}%
Now, the sets $({Q_{i}\cap Q_{i-1})\setminus Q_{i-2}}$ and ${%
Q_{i-1}\setminus \left( {Q_{i}\cup Q_{i-2}}\right) }$ partition the set ${%
Q_{i-1}\setminus Q_{i-2}}$, which has $a_{i-1}-a_{i-2}$ nodes, due to (\ref%
{eqA1}). Thus, it follows from (\ref{eqA2}) that%
\begin{equation}
\left\vert {Q_{i-1}\setminus \left( {Q_{i}\cup Q_{i-2}}\right) }\right\vert
=a_{i-1}-a{_{i-2}}-\sum\nolimits_{\ell =a_{i-2}+1}^{a_{i-1}}{H\left( {\ell
+x_{\ell }-a_{i}}\right) }  \label{eqA3}
\end{equation}%
In order to compute the value $\left\vert {Q_{i}\cap Q_{i-2}}\right\vert $,
we have to compute how many of the $1^{st},2^{nd},...,a_{i-2}^{th}$ diagonal
elements belong simultaneously to $Q_{i-2}$ and to $Q_{i}$. For $1\leq \ell
\leq a_{i-2}$, the $\ell ^{th}$ one belongs to $Q_{i}$ exactly when its
chain of $1$'s reaches the $a_{i}^{th}$ row, i.e. exactly when $\ell
+x_{\ell }\geq a_{i}$, or equivalently $H\left( {\ell +x_{\ell }-a_{i}}%
\right) =1$. In this case, if $\ell \neq a_{i-2}$, then its chain reaches
also the $a_{i-2}^{th}$ row, which means that it belongs also to $Q_{i-2}$,
while the $a_{i-2}^{th}$ one belongs always to $Q_{i-2}$. It follows that 
\begin{equation}
\left\vert {Q_{i}\cap Q_{i-2}}\right\vert =\sum\nolimits_{\ell =1}^{a_{i-2}}{%
H\left( {\ell +x_{\ell }-a_{i}}\right) }  \label{eqA4}
\end{equation}%
Similarly, in order to compute the value $\left\vert ({Q_{i-1}\cap
Q_{i-2})\setminus Q}_{i}\right\vert $, we have to compute how many of the $%
1^{st},2^{nd},...,a_{i-2}^{th}$ diagonal elements belong simultaneously to $%
Q_{i-1}$ and to $Q_{i-2}$ but not to $Q_{i}$. For $1\leq \ell \leq a_{i-2}$,
the $\ell ^{th}$ one belongs to $Q_{i-1}$ exactly when $H\left( {\ell
+x_{\ell }-a_{i-1}}\right) =1$. In this case it belongs also to $Q_{i-2}$,
since $a_{i-2}<a_{i-1}$. Further, it does not belong to $Q_{i}$ exactly when 
$\ell +x_{\ell }<a_{i}$, or equivalently ${H\left( a_{i}-{\ell -x_{\ell }-1}%
\right) }=1$. It follows that 
\begin{equation}
\left\vert ({Q_{i-1}\cap Q_{i-2})\setminus Q}_{i}\right\vert
=\sum\nolimits_{\ell =1}^{a_{i-2}}{H\left( {\ell +x_{\ell }-a_{i-1}}\right)
\cdot H\left( a_{i}-{\ell -x_{\ell }-1}\right) }  \label{eqA5}
\end{equation}%
Finally, the complementary part in $G_{i}$ of the sets in (\ref{eqA1})-(\ref%
{eqA5}) has 
\begin{equation}
a_{i-2}-\sum\nolimits_{\ell =1}^{a_{i-2}}{H\left( {\ell +x_{\ell }-a_{i}}%
\right) }-\sum\nolimits_{\ell =1}^{a_{i-2}}{H\left( {\ell +x_{\ell }-a_{i-1}}%
\right) \cdot H\left( a_{i}-{\ell -x_{\ell }-1}\right) }  \label{eqA6}
\end{equation}%
nodes, since $G_{i}$ has overall $a_{i}$ nodes.
\end{appendix}

\end{document}